\documentstyle{mn}

\input{epsf}

\def\mpc {h^{-1} {\rm Mpc}}

\def\and  {{\it {et al.} }}

\newcommand{\etal}{{\it et al.\ }}
\newcommand{\xiav}{\bar{\xi}}

\newcommand{\avg}[1]{\langle{#1}\rangle}

\newcommand{\ltsima}{$\; \buildrel < \over \sim \;$}
\newcommand{\lsim}{\lower.5ex\hbox{\ltsima}}
\newcommand{\gtsima}{$\; \buildrel > \over \sim \;$}
\newcommand{\gsim}{\lower.5ex\hbox{\gtsima}}

\begin{document}

\title[Determining Bias with Cumulant Correlators]{
Determining Bias with Cumulant Correlators}

\author[I.Szapudi]{Istv\'an Szapudi$$
%\altaffilmark{1}}
%\affil{Fermi National Accelerator Laboratory}
%\affil{Theoretical Astrophysics Group}
%\affil{Batavia, IL 60510}
%\altaffiltext{1}{E-mail:\ szapudi@astro1.fnal.gov}
\\
\\
University of Durham, Department of Physics 
 South Road,  Durham DH1 3LE, United Kingdom               
\\
}

\maketitle 
 
\begin{abstract}

The first non-trivial cumulant correlator
of the galaxy density field  $Q_{21}$ is examined from the point
of view of biasing. It is shown that to leading order it depends on 
two biasing parameters $b$, and $b_2$, and on $q_{21}$, the underlying
cumulant correlator of the mass. As the skewness $Q_3$
has analogous properties, 
the slope of the correlation function $-\gamma$,
$Q_3$, and $Q_{21}$ uniquely determine the bias parameter
on a particular scale to be $b = \gamma/6(Q_{21}-Q_3)$,
when working in the context of gravitational instability 
with Gaussian initial conditions.
Thus on large scales,  easily accessible with the future Sloan Digital
Sky Survey and the 2 Degree Field Survey, 
it will be possible to extract $b$, and $b_2$
from simple counts in cells measurements. Moreover, the higher
order cumulants, $Q_N$, successively determine the higher order
biasing parameters. From these it is possible 
to predict higher order cumulant
correlators as well. Comparing the predictions with the measurements
will provide internal consistency checks on the validity
of the assumptions in the theory, most notably
perturbation theory of the growth of fluctuations by gravity and 
Gaussian initial conditions. Since the method is insensitive to
$\Omega$, it can be successfully combined with results from
velocity fields, which determine $\Omega^{0.6}/b$, to measure
the total density parameter in the universe.

\end{abstract}


\section{Introduction}

Galaxies are likely to be biased tracers of the underlying
density field, since surveys with different selection properties
are correlated  differently, i.e. they are biased with respect to each other. 
The amplitude of the two point correlation function is ambiguous and 
cannot be compared directly with measurements of the cosmic microwave 
background fluctuations, or dark matter simulations. 
Although the galaxy density field can be a complicated function of
the underlying mass field, on large scales
the fluctuations are small and
only the leading order contributions are important.  Then
the two-point function of galaxies becomes $b^2$ times the
underlying two-point correlation function, where $b = b_1$
is the first coefficient in the Taylor expansion of the
general (local) bias function, the galaxy density field as 
function of the underlying mass field.

Higher order statistics can be used to alleviate this ambiguity
by constraining the unknown value of $b$.
The cumulants of the galaxy field are related to
the cumulants of the underlying mass field and the biasing
parameters (Fry and Gazta\~naga 1993, hereafter FG93). For instance the 
skewness of the galaxy field was shown to be
$Q_3 = q_3/b+b_2/b^2$, where $b_2$ is the second order coefficient
in the expansion of the bias function, and $q_3$, and $Q_3$ are
the skewness of the mass and galaxy field, respectively. 
The slope of the two-point function determines
the theoretical $q_3$ using perturbation theory
\cite{peebles80,f94,bern92,jbc93,bern94}. 
Unfortunately $Q_3$ alone cannot yield both bias parameters, 
therefore additional information is needed. 
Note that adding $Q_4$ does not help, since to leading order it depends
on an additional bias parameter, $b_3$, and at higher 
orders the situation is exactly analogous. 
Several attempts have been made
in the literature to supply the missing piece of information
and to constrain the bias parameter 
by using the bi-spectrum 
\cite{f94,mvh97,scffhm98},
configuration dependence of the three-point function \cite{fg98},
and the skewness in elongated cells \cite{ssf98} 
This {\em Letter}
proposes a simple, robust alternative using the first non-trivial
cumulant correlator \cite{ss97a} to provide the needed additional 
constraint.  This technique yields $b$ via counts in cells
measurements, thus it is possibly one of the simplest and most unique
methods ever put forward.
Note that there are alternatives besides higher
order statistics for the determination of the bias, most
importantly the use of velocity fields
\cite{wsdk97}, which, however, determine
the combination of $\Omega^{0.6}/b$. 
The simple method proposed here is insensitive to the
cosmological parameter $\Omega$. Thus in combination with results
from velocity flows the total density in the universe 
can be determined.

Section \S2 introduces the theory of biasing for cumulant
correlators, \S3 details how it can be applied to measure
the bias parameters \S4 discusses practicality, 
possible extensions,  and limitations of the results.

\section{Biasing of Cumulant Correlators}

Let us consider the moments and joint moments of the 
(smoothed) fluctuation field
$\delta$, and denote the connected moments of the
field with $\avg{}_c$.
The cumulants,  $q_N$, can be defined by the moments as
$\xiav = \avg{\delta^2}$, and 
$\avg{\delta^N}_c = q_N N^{N-2}\xiav^{N-1}\,\, (N \ge 3)$.  Similarly
the cumulant correlators are defined
through the joint moments $\xi_l = \avg{\delta_1 \delta_2}$ and
$\avg{\delta_1^N\delta_2^M}_c = q_{NM} N^{N-1}M^{M-1}\xiav^{M+N-2}\xi_l 
\,\, (N+M \ge 3)$.
 The former is a well established tool for characterization
of the non-Gaussianity of the galaxy distribution
\cite{peebles80}, while the latter were recently introduced
by \cite{ss97a} based on factorial moment correlators 
\cite{ssb92,mss92,sdes95}.

If the galaxy density field is a general function of the underlying
density field $\delta_g = f(\delta)$, similar quantities can
be defined for the galaxies. Note that this {\em Letter}
uses continuum limit throughout, even though galaxies constitute
a discrete process. This discreteness can be simply taken into
account by the use of factorial moments \cite{ss93a} under
the assumption of infinitesimal Poisson sampling \cite{peebles80}, and will
not be mentioned further. On large scales, where the mass
fluctuations are small, it makes sense to Taylor expand the
galaxy density field in terms of the mass field
(bias function) $f(\delta) = \sum_{k\ge 0} b_k\delta^k/k!$, where $b_k$
are called the $k$th order bias parameters. FG93 have
shown that to leading order it is possible to calculate
the cumulants of the galaxy field in terms of cumulants of the underlying
density field, and the bias parameters. Next their results are
recapitulated briefly. For $N \le 3$ the connected moments of
the galaxy field are
\begin{eqnarray}
   \xiav_g  =& b^2 \xiav \\
    Q_3     =& q_3/b+b_2/b^2 \nonumber,
\end{eqnarray}
where $b_1 = b$, the usual bias parameter for simplicity.
The bias can be systematically calculated
from the generating function $\log\avg{e^{f(\delta) x}}$, and
 explicit results up to sixth order can be found in FG93. 
The parameter $b_0$ is irrelevant for the
higher order cumulants, it can be fixed by
requiring $\avg{f(\delta)} = 0$. The hierarchy is approximately
inherited by the biased field to lowest non-vanishing order,
even for multipoint  estimators.
Note that several other general calculations were done in the
past for the correlations of biased fields, 
among others \cite{k84,bbks86,gw86,mlb86,s88,s94,m95}.

Analogously to FG93, biasing of the cumulant correlators
can be calculated from the generating function 
$\log\avg{e^{f(\delta_1)x + f(\delta_2)y}}$ \cite{ss93a}. 
It is a trivial although tedious matter to derive
explicit formulae from it.
Here only those results are shown which are needed for the applications
in the next section,
\begin{eqnarray}
   \xi_{l,g} =& b^2 \xi_l \\
   Q_{21}    =& q_{21}/b+b_2/b^2 \nonumber,
\end{eqnarray}
where the notation is analogous to the previous equation. 
Note that the dependence on the bias parameters is extremely
simple and exactly analogous to the corresponding results for $Q_3$.
This facilitates the solution of the equations in the next section.
The higher order results will be shown explicitly elsewhere. Note that
the generating function can be trivially generalized
to higher than two-point estimators as well.

\section{Measuring the Bias Parameter}

For models of structure formation where
Gaussian initial conditions were amplified via gravity,
the above formulae can be combined with perturbation
theory results in the weakly non-linear
regime \cite{peebles80,f84,bern94,bern96}
\begin{eqnarray}
   q_3 =& 34/21-\gamma/3 \\
   q_{21} =& 34/21-\gamma/6 \nonumber,
\end{eqnarray}
where $-\gamma$ is the slope of the mass correlation function at
the particular scale considered. Note that if the bias
is a slowly varying function of scale, the slope of the mass
and galaxy correlation functions are the same on a given
scale. This plausible condition is assumed for the rest
of the paper. The above equations were
found to be good approximations in both galaxy catalogs
and simulations  
\cite{peebles80,cf91,gaz92,bouchet93,gaz94,cbh95,smn96,sqsl98}.

Thus given $\gamma$,
$Q_{3}$, and $Q_{21}$, the equations can be solved for the
bias parameters $b,b_2$. The result 
is
\begin{equation}
  b   = \frac{\gamma}{6(Q_{21}-Q_{3})} \\
\end{equation}
\begin{equation}
  b_2 = \frac{\gamma Q_{21}(14\gamma-68)+\gamma Q_3(68-7\gamma)}
         {252(Q_{21}-Q_3)^2}\nonumber. 
\end{equation}
These equations constitute the main result of this
{\em Letter}. They become degenerate when $q_{21} = q_3$, i.e. $\gamma = 0$.
For typical CDM spectra and for $b \ne 0$ the solution is 
well behaved on large scales, where the assumptions of this calculation
hold. The above solution provides a simple and unique way of determining the
bias parameter $b$. Note that $Q_{21}$ and the $Q_N$'s
up to a certain order  fully resolve the degeneracy, 
as the $Q_N$'s depend on $N-1$ bias parameters, and $\gamma$, 
thus the $b_{N-1}$'s can be successively
calculated. From these further cumulant
correlators can be predicted and compared with measurements,
providing a set of consistency checks for the theory.

\section{Discussion}

The previous section showed, that if $-\gamma$, the slope of the
correlation function, $Q_3$, the third order cumulant, and
$Q_{21}$ the first non-trivial cumulant correlator are measured,
the bias parameter $b$ can be calculated. Next the practicality
of such a measurement in the near future will be reviewed.

According to \cite{css98}, the third and fourth moments of the galaxy
distribution will be measured with less than a few \%, and
$10$\% error, respectively, 
in the Sloan Digital Sky Survey (SDSS) on scales between
$1-50\mpc$. Errors of the the same order can be expected
for the future 2 Degree Field Survey (2dF), perhaps in
a slightly narrower dynamic range of scales because of
its smaller size. Similar accuracy can be achieved 
for $Q_3$ and $Q_{21}$, and even better
accuracy for the correlation function on these scales.
The previous considerations require that $\delta << 1$
both because of the Taylor
expansion of the bias function, and reliance on perturbation theory.  
This still leaves
quite a large dynamic range, roughly $10-50\mpc$, in which
the bias parameters $b$, and $b_3$, and their scale dependence can be determined
accurately. Similarly, at higher orders
using $Q_N$, $b_{N-1}$ can be calculated successively. Thus on large scales
the bias will be accurately measured facilitating
the comparison with the fluctuations in the cosmic microwave
background and $N$-body simulations.
Note that the assumption of large scales is
important for both the applicability of the perturbation
theory and Taylor expansion.
On small scales, where the hierarchical assumption applies,
i.e. the three-point function
$\zeta = Q_3(\xi_1\xi_2+\xi_2\xi_3+\xi_3\xi_1)$ for
$N = 3$ \cite{peebles80}, 
$Q_{21} - Q_3 = \xi_l^2/2 $, as found in the APM
\cite{ss97a}.  This leads to an asymptotically small bias formally
$b = \gamma/3\xi_l^2$, because the first term
in the Taylor expansion does not describe the bias
accurately when the variance is large.

Besides systematic errors, which are expected to be quite small
for the SDSS and 2dF, redshift distortions could pose a problem. 
For the two-point function the theory of redshift distortions
is well known \cite{k87}, and can be corrected for.
For $Q_3$ the effects
of redshift distortions were calculated by \cite{hbcj95}, and 
they found it to be negligible on large scales; simulations 
\cite{scfh98} confirm this on scales when the variance is small, which is
anyway the limit of applicability of the theory described
above. It is safe to assume that the same holds for $Q_{21}$
as well, since it is a third order cumulant with slightly
different smoothing than $Q_3$.
Thus redshift distortions do not constitute a real 
problem for applying the theory of the previous section. 
The light cone effect is even less of a problem than redshift distortions: 
higher order moments of the distribution can be corrected for it, 
and in any case it is not
expected to be important for the next generation of wide
field surveys where the median
redshift is $z \simeq 0.2$ \cite{mss98}.

Practicality of measuring these key quantities was demonstrated
by several authors: correlation functions, and cumulants
were estimated in numerous galaxy catalogs and simulations
\cite{peebles80,gaz92,ssb92,mss92,bouchet93,gaz94,sdes95,cbh95,smn96}, 
cumulant correlators
were determined in the APM catalog \cite{ss97a}, and recently
in simulations \cite{mm98}. In fact, the slope $-\gamma$ in 
the previous equations is precisely the slope of the second
moment of fluctuations in the same cells in which the cumulants
and cumulant correlators are measured. Therefore in practice, only
counts in cells and bicounts (or joint counts) in pairs of cells 
have to be estimated for obtaining
all required quantities.  
Thus, there is no doubt that this simple method will give a 
unique determination of the bias on large scales from the SDSS, 
and from the 2dF.

There are several further developments possible for the simple theory
of the previous sections. First of all, it is entirely trivial to calculate
the effects of bias on higher order cumulant correlators,
given the generating function of \S1. Similarly, the generating
function generalizes easily for multipoint statistics.
Projected catalogs can provide another way of estimating the needed three
quantities, although deprojection in the weakly non-linear
regime is far from trivial matter \cite{bern95,gb97}.
Smaller scales could
be penetrated as well by using either extended perturbation
theory \cite{cbbh97}, or second order perturbation theory
\cite{sf96a,sf96b} in combination with expanding the bias correction
further than leading order. Note that smoothing in 
this theory precedes biasing as in FG93.
While these two operations do not commute, according to
David Weinberg (1998, private
communication) the effect of this appears to be negligible
in simple models on large scales, although it could lead
to problems for some exotic biasing schemes. 
Based on the group properties of the bias function
(FG94) it is possible to apply Equations 1-2, and
their above mentioned possible generalizations,  for
biasing between clusters and galaxies, or 
between different types of galaxies
\cite{k84,bs92,ss93b,bern96} to predict
the enhancement of the two-point correlation function
on large scales from counts in cells. The mentioned group
property enables more
then one step as well, i.e. between mass to galaxies to clusters, 
etc.  Note that all the 
previous considerations were based on local bias
models. It is possible to construct more general
models which are stochastic and/or non-local in nature.
These possible improvements and generalizations, 
as well as a full scale demonstration of
the technique on mock galaxy catalogs \cite{chwf98}
with different biases will be presented elsewhere. 

In summary, a novel method has been presented to determine
the bias parameter $b$, as well as the higher order
bias parameters. It is simple to use and will become
a practical possibility in the near future when the
new generation of wide field surveys,
the SDSS and 2dF, come on line. The technique
involves the estimation of (bi)counts in cells, and from them
the slope of the correlation
function $-\gamma$, and two third order moments: $Q_3$, and $Q_{21}$.
The bias parameter can be obtained
explicitly as $b = \gamma/6(Q_{21}-Q_3)$ in the
context of gravitational instability with Gaussian initial
conditions. It was demonstrated that
these quantities can be measured in the near future, and
$N > 3$ connected moments will provide additional constraints
as internal consistency check. Thus
a unique determination of the bias will be a realistic
possibility with higher order statistics. Since velocity measurements
yield $\Omega^{0.6}/b$, this method can be combined with bulk flow
measurements to yield $\Omega$ as well. The general theory
presented here can be used to describe bias between
different species of galaxies and clusters as well.

%\begin{figure}
%\centering
%\centerline{\epsfysize=9.truecm
%\epsfbox{s3.ps}}
%\caption[junk]{Same as in Figure \ref{sigma} for the 
%the hierarchical skewness $s_3=w_3/w_2^2$.
%The misalignment of the open and solid  squares on scales $gt 0.5$ degrees
%is the  result of edge effects, as both correspond to smaller surveys.}
%\label{s3}
%\end{figure}

\bigskip

{\bf Acknowledgments}

It is a pleasure to acknowledge stimulating discussions
with Carlos Frenk and Alex Szalay. 
The author was supported by the PPARC rolling grant for 
Extragalactic Astronomy and Cosmology at Durham.

\def\apj { ApJ}
\def\aap {A \& A}
\def\ajs{ ApJS}
\def\apjs{ ApJS}
\def\mnras { MNRAS}
\def\apjl { Ap. J. Let.}

\end{document}